\documentclass[english]{revtex4-2}
\usepackage[T1]{fontenc}
\usepackage[latin9]{inputenc}
\usepackage{amsmath}
\usepackage{amssymb}

\makeatletter
\usepackage{babel}
\def\CM{\mathrm{\scriptscriptstyle CM}}

\makeatother

\usepackage{babel}
\begin{document}
\title{An observable for the separation distance of two photons}
\author{Scott E. Hoffmann}
\address{School of Mathematics and Physics,~~\\
 The University of Queensland,~~\\
 Brisbane, QLD 4072~~\\
 Australia}
\email{scott.hoffmann@uqconnect.edu.au}

\begin{abstract}
Motivated by an intention to apply partial wave analysis to systems
containing photons, we construct eigenvectors of the distance of separation
between two photons. A choice is made that makes the case of two photons
most closely resemble the case of two massive particles. The treatment
is relativistic. An Hermitian separation observable can then be defined,
unlike a position operator for a single photon, which cannot be defined.
We test these results on a model system of two photons meant to describe
the initial or final state of a scattering experiment. We find that
the separation probability density gives a meaningful picture of localization.
\end{abstract}
\maketitle

\section{Introduction}

I has long been known that it is not possible to construct point-localized
state vectors for individual photons that satisfy the criteria of
\citet{Newton1949}. We consider that these criteria capture the essence
of what it means to have a point-localized state vector in quantum
mechanics, so we have no intent, in this paper, of challenging those
criteria. Some authors (\citet{Kiessling2018,Bialynicki-Birula2009,Hawton1999a,Pryce1948})
claim to have produced a position operator for the photon, but these
are defined to act on proposed wavefunctions rather than on the basis
$\{|\,k,\lambda\,\rangle\}$ of momentum-helicity eigenvectors. In
this paper, the observable we introduce has matrix elements between
direct products of photon momentum-helicity eigenvectors.

The reason for this negative result is not the masslessness of the
photon, as point-localized state vectors for a hypothetical massless,
spin-0 particle could be constructed. Instead, the reason is the limited
helicity spectrum, $\lambda=\pm1,$ of the photon (\citet{Wightman1962}).
Because of the way eigenvectors of helicity rotate, it is not possible
to form an irreducible representation of rotations about the localization
point without having an extra, $\lambda=0,$ state.

Yet it is clear from physical results that photons can be partially
localized in wavepacket states. From a theoretical perspective, we
would like to perform partial wave analysis on, for example, two-particle
states including one or two photons. To find the phase shifts requires
finding the asymptotic form, as the separation grows without bound,
of amplitudes invlving basis vectors with particle separation as an
eigenvalue.

Thus we are led to find eigenvectors of photon separation rather than
individual positions. We will do this using eigenvectors of total
four-momentum ($P$), total centre of mass (CM) frame angular momentum
(quantum numbers $J$ and $M$) and two CM-frame helicities ($\lambda_{1}^{\CM}$
and $\lambda_{2}^{\CM}$). In these basis vectors, the rotation properties
of the helicity-carrying photons are ``hidden''. The state vectors
rotate only according to the $\boldsymbol{J}^{2}=J(J+1)$ representations
in the CM frame.

A reasonable choice, introduced as a postulate, is made to quantify
these concepts. Then the consequences of this choice are calculated
for a model system. The result is a useful measure of separation distance
for two photons, indistinguishable, for the case considered, from
the result for two massive particles.

Throughout this paper, we use Heaviside-Lorentz units, in which $\hbar=c=\epsilon_{0}=\mu_{0}=1.$

\section{The separation of two massive particles}

The following construction can only be done for two massive particles,
and provides a comparison with the one- or two-photon cases. We want
to define separation eigenvectors in terms of eigenvectors of total
four-momentum, $P\rightarrow P^{\mu}$ for $\mu=0,1,2,3$ and total
CM-frame angular momentum (quantum numbers $J,M$), since we can construct
similar eigenvectors for the one- and two-photon cases, allowing that
comparison.

We consider two different individual particle translations applied
to a free two-particle eigenvector of the individual momenta and the
rest-frame spin $z$-components (eigenvalues $\mu_{1}$ and $\mu_{2}$
for spins $s_{1}$ and $s_{2},$ respectively):
\begin{equation}
U(T_{1}(+\boldsymbol{r}/2))\,U(T_{2}(-\boldsymbol{r}/2))\,|\,\frac{1}{2}\boldsymbol{P}+\boldsymbol{p},\mu_{1},\frac{1}{2}\boldsymbol{P}-\boldsymbol{p},\mu_{2};f\,\rangle=|\,\frac{1}{2}\boldsymbol{P}+\boldsymbol{p},\mu_{1},\frac{1}{2}\boldsymbol{P}-\boldsymbol{p},\mu_{2}\,\rangle\,e^{-i\boldsymbol{p}\cdot\boldsymbol{r}}.\label{eq:2.1}
\end{equation}
(We are using a ``noncovariant'' orthonormalization for the basis
vectors (\citet{Fong1968}). See eq. (\ref{eq:2.6}) below.) This
transformation changes the separation of the two particles by $\boldsymbol{r}$
but leaves the average position unchanged. So we define simultaneous
eigenvectors of separation in the CM frame (where $\boldsymbol{P}=0$)
and zero total three-momentum as
\begin{equation}
|\,\boldsymbol{r}^{\CM},\boldsymbol{0},\mu_{1},\mu_{2};f\,\rangle=\int\frac{d^{3}p}{(2\pi)^{\frac{3}{2}}}\,|\,+\boldsymbol{p},\mu_{1},-\boldsymbol{p},\mu_{2};f\,\rangle\,e^{-i\boldsymbol{p}\cdot\boldsymbol{r}^{\CM}},\label{eq:2.2}
\end{equation}
noting that
\begin{equation}
\hat{\boldsymbol{r}}^{\CM}=i\frac{\partial}{\partial\boldsymbol{p}}=i\frac{\partial}{\partial\boldsymbol{p}_{1}}-i\frac{\partial}{\partial\boldsymbol{p}_{2}}\label{eq:2.3}
\end{equation}
and $\hat{\boldsymbol{P}}$ commute and both commute with the spin
operators.

We form eigenvectors of the magnitude of the separation distance and
of orbital angular momentum in the CM frame,
\begin{equation}
|\,r^{\CM},l,m,\boldsymbol{0},\mu_{1},\mu_{2};f\,\rangle=\int d^{2}\hat{r}^{\CM}\,|\,\boldsymbol{r}^{\CM},\boldsymbol{0},\mu_{1},\mu_{2};f\,\rangle\,r^{\CM}\,Y_{lm}(\hat{\boldsymbol{r}}^{\CM}),\label{eq:2.4}
\end{equation}
where the argument of the spherical harmonic is here a unit vector,
not an operator. We note that $\hat{\boldsymbol{L}}=\hat{\boldsymbol{r}}\times\hat{\boldsymbol{p}},$
defined in the CM frame, also commutes with $\hat{\boldsymbol{P}}$
and the spin operators.

Note that these basis vectors will have calculable but very complicated
Lorentz transformation properties, as with any position eigenvectors
(\citet{Newton1949}). For our purposes it will suffice to ignore
the total three-momentum label in what follows.

Then we combine the spins using Clebsch-Gordan coefficients (\citet{Messiah1961}),
producing a set of representations with quantum numbers $S$ in the
range $|s_{1}-s_{2}|\leq S\leq s_{1}+s_{2}.$ Finally we combine this
added spin angular momentum with the orbital angular momentum to find
\begin{multline}
|\,r^{\CM},J,M,\boldsymbol{0},l,S;f\,\rangle\\
=\int d^{2}\hat{r}^{\CM}\,|\,\boldsymbol{r}^{\CM},\boldsymbol{0},\mu_{1},\mu_{2};f\,\rangle\,r^{\CM}\,Y_{lm}(\hat{\boldsymbol{r}}^{\CM})\,\langle\,s_{1}\,s_{2}\,\mu_{1}\,\mu_{2}\,|\,S\,M_{S}\,\rangle\langle\,l\,S\,m\,M_{S}\,|\,J\,M\,\rangle\quad\mathrm{for}\ J\geq|s_{1}-s_{2}|,\label{eq:2.5}
\end{multline}
with summations implied over $\mu_{1},\mu_{2},m$ and $M_{S}.$

If we choose the orthonormalization
\begin{equation}
\langle\,+\boldsymbol{p}_{a},\mu_{1a},-\boldsymbol{p}_{a},\mu_{2a};f\,|\,\boldsymbol{p}_{b},\mu_{1b},-\boldsymbol{p}_{b},\mu_{2b};f\,\rangle=\delta_{\mu_{1a}\mu_{1b}}\delta_{\mu_{2a}\mu_{2b}}\,\delta^{3}(\boldsymbol{p}_{a}-\boldsymbol{p}_{b}),\label{eq:2.6}
\end{equation}
then we find
\begin{equation}
\langle\,r_{a}^{\CM},J_{a},M_{a},l_{a},S_{a};f\,|\,r_{b}^{\CM},J_{b},M_{b},l_{b},S_{b};f\,\rangle=\delta_{J_{a}J_{b}}\delta_{M_{a}M_{b}}\delta_{l_{a}l_{b}}\delta_{S_{a}S_{b}}\,\delta(r_{a}^{\CM}-r_{b}^{\CM}),\label{eq:2.7}
\end{equation}
confirming $r^{\CM}$ an a Hermitian observable. Note that if we were
dealing with identical bosons or fermions, symmetrization or antisymmetrization,
respectively, could be applied.

A similar construction could be used to form eigenvectors of momentum
magnitude and orbital angular momentum in the $\boldsymbol{P}=0$
CM frame, starting with
\begin{equation}
|\,p^{\CM},l,m,\mu_{1},\mu_{2};f\,\rangle=\int d^{2}\hat{p}\,|\,\boldsymbol{p},\mu_{1},-\boldsymbol{p},\mu_{2};f\,\rangle\,p\,Y_{lm}(\hat{\boldsymbol{p}}).\label{eq:2.8}
\end{equation}
Then we find the relation between eigenvectors of momentum magnitude
and eigenvectors of separation magnitude:
\begin{equation}
\langle\,r_{b}^{\CM},J_{b},M_{b},l_{b},S_{b};f\,|\,p^{\CM},J_{b},M_{b},l_{b},S_{b};f\,\rangle=\delta_{J_{a}J_{b}}\delta_{M_{a}M_{b}}\delta_{l_{a}l_{b}}\delta_{S_{a}S_{b}}\,\sqrt{\frac{2}{\pi}}\,p^{\CM}r^{\CM}\,j_{l_{a}}(p^{\CM}r^{\CM}),\label{eq:2.9}
\end{equation}
in terms of the free spherical waves, $y_{l}^{(f)}(r,p)=\sqrt{\frac{2}{\pi}}\,pr\,j_{l}(pr),$
familiar from partial wave analysis of the radial, \textit{nonrelativistic},
Schrödinger equation (\citet{Messiah1961}). But nowhere in its derivation
here did we specify the nonrelativistic dependence of energy on momentum.
They obey the differential equation
\begin{equation}
\{-\frac{d^{2}}{dr^{2}}+\frac{l(l+1)}{r^{2}}\}\,y_{l}^{(f)}(r,p)=p^{2}y_{l}^{(f)}(r,p).\label{eq:2.10}
\end{equation}
There is no mass in this equation. It is not saying $p^{2}/2m$ is
the energy for some mass, $m.$ It is merely saying that the square
of the individual-particle momenta in the CM frame is $p^{2}$ for
all partial waves with $l\geq0.$ As an equation in the centre of
mass frame, it is consistent with special relativity.

\section{The separation of two photons}

For a two-photon state vector, it is not possible to define orbital
and spin angular momenta separately, so the previous construction
is not possible. Instead, we use the helicity formalism (\citet{Jacob1959,Macfarlane1962}),
which produces basis vectors $|\,P,J,M,\lambda_{1}^{\CM},\lambda_{2}^{\CM};f:\gamma\gamma\,\rangle.$
In particular, in the CM frame,
\begin{multline}
|\,k^{\CM},\boldsymbol{0},J,M,\lambda_{1}^{\CM},\lambda_{2}^{\CM};f:\gamma\gamma\,\rangle\\
=\int d^{2}\hat{\boldsymbol{k}}^{\CM}\,|\,+\boldsymbol{k}^{\CM},\lambda_{1}^{\CM},-\boldsymbol{k}^{\CM},\lambda_{2}^{\CM};f:\gamma\gamma\,\rangle e^{-i\lambda_{2}^{\CM}(2\varphi_{1}^{\CM}+\pi)}\,\sqrt{\frac{2J+1}{4\pi}}\,\mathcal{R}_{M\lambda_{1}^{\CM}-\lambda_{2}^{\CM}}^{(J)*}[\hat{\boldsymbol{k}}^{\CM}],\label{eq:3.1}
\end{multline}
for $J\geq|\lambda_{1}^{\CM}-\lambda_{2}^{\CM}|,$ where $\theta_{1}^{\CM}$
and $\varphi_{1}^{\CM}$ are the spherical polar angles of particle
1 and $\mathcal{R}_{M\lambda_{1}^{\CM}-\lambda_{2}^{\CM}}^{(J)*}[\hat{\boldsymbol{k}}^{\CM}]$
are matrix elements of the unitary representation of the rotation
\begin{equation}
R_{0}(\hat{\boldsymbol{k}}^{\CM})=R_{z}(+\varphi_{1}^{\CM})R_{y}(\theta_{1}^{\CM})R_{z}(-\varphi_{1}^{\CM}).\label{eq:3.2}
\end{equation}
The eigenvalue of total energy is $2k^{\CM};$ we merely renormalized
to eigenvalue $k^{\CM}.$ Again we factor out the dependence on total
three-momentum in what follows.

Our physical expectation is that the separation probability density
has no dependence on spin or helicity, at least for large particle
separations. Thus we propose a definition that makes the case of two
photons most closely resemble the case of two massive particles (for
$r^{\CM}>0$),
\begin{multline}
\langle\,r^{\CM},J_{a},M_{a},\lambda_{1a}^{\CM},\lambda_{2a}^{\CM};f:\gamma\gamma\,|\,k^{\CM},J_{b},M_{b},\lambda_{1b}^{\CM},\lambda_{2b}^{\CM};f:\gamma\gamma\,\rangle\\
=\delta_{J_{a}J_{b}}\delta_{M_{a}M_{b}}\delta_{\lambda_{1a}^{\CM}\lambda_{1b}^{\CM}}\delta_{\lambda_{2a}^{\CM}\lambda_{2b}^{\CM}}\,\sqrt{\frac{2}{\pi}}\,k^{\CM}r^{\CM}\,j_{\ell}(k^{\CM}r^{\CM}),\label{eq:3.3}
\end{multline}
which defines the separation eigenvectors, in abbreviated form, as
\begin{equation}
|\,r^{\CM},Q;f:\gamma\gamma\,\rangle=\int_{0}^{\infty}dk^{\CM}\,|\,k^{\CM},Q;f:\gamma\gamma\,\rangle\,\sqrt{\frac{2}{\pi}}\,k^{\CM}r^{\CM}\,j_{\ell}(k^{\CM}r^{\CM}),\label{eq:3.4}
\end{equation}
guaranteeing orthonormality, since
\begin{equation}
\int_{0}^{\infty}dk\,\sqrt{\frac{2}{\pi}}\,kr\,j_{\ell}(kr_{a})\,\sqrt{\frac{2}{\pi}}\,kr\,j_{\ell}(kr_{b})=\delta(r_{a}-r_{b}).\label{eq:3.5}
\end{equation}
Here $Q=J,M,\lambda_{1}^{\CM},\lambda_{2}^{\CM}$ represents the other
quantum numbers.

We are left with the question of what to choose for $\ell.$ No orbital
angular momentum quantum number appears in these expressions. We argue
that a suitable choice is
\begin{equation}
\ell=\ell(J,\lambda_{1}^{\CM},\lambda_{2}^{\CM})=J-|\lambda_{1}^{\CM}-\lambda_{2}^{\CM}|.\label{eq:3.6}
\end{equation}
It is always integral and its spectrum, from the condition attached
to eq. (\ref{eq:3.1}), is $\ell=0,1,2,\dots$.

The observable with eigenvalues $r^{\CM}$ is then given by
\begin{equation}
\hat{r}^{\CM}=\int_{0}^{\infty}d\rho^{\CM}\sum_{Q}\,|\,\rho^{\CM},Q;f:\gamma\gamma\,\rangle\,r\,\langle\,\rho^{\CM},Q;f:\gamma\gamma\,|.\label{eq:3.7}
\end{equation}

Note that the case of one photon and one massive particle could be
easily treated using this method.

\section{Calculation of a separation probability density}

From our experience with the massive case, and use of the measure
of localization introduced in \citet{Hoffmann2020a}, we expect that
the state vector (not yet symmetrized)
\begin{equation}
|\,\psi(\boldsymbol{k}_{0},\boldsymbol{R},\lambda_{1},\lambda_{2})\,\rangle=\int d^{3}k_{1}\int d^{3}k_{2}\,|\,\boldsymbol{k}_{1},\lambda_{1},\boldsymbol{k}_{2},\lambda_{2}\,\rangle\,\frac{e^{-|\boldsymbol{k}_{1}-\boldsymbol{k}_{0}|^{2}/4\sigma_{k}^{2}}}{(2\pi\sigma_{k}^{2})^{\frac{3}{4}}}\,e^{+ik_{1z}R/2}\,\frac{e^{-|\boldsymbol{k}_{2}+\boldsymbol{k}_{0}|^{2}/4\sigma_{k}^{2}}}{(2\pi\sigma_{k}^{2})^{\frac{3}{4}}}\,e^{-ik_{2z}R/2}\label{eq:4.1}
\end{equation}
represents one photon with average momentum $\boldsymbol{k}_{0}=+k_{0}\hat{\boldsymbol{z}}$
($k_{0}>0$) and a small spread in momentum, chosen so that $\sigma_{k}\ll k_{0},$
partially localized at average position $-R\hat{\boldsymbol{z}}/2$
with large spatial width, $\sigma_{r}$ (with $\sigma_{r}\sigma_{k}=\frac{1}{2}$),
and helicity $\lambda_{1},$ and another photon with average momentum
$-\boldsymbol{k}_{0}$ partially localized around $+R\hat{\boldsymbol{z}}/2$
with helicity $\lambda_{2},$ at time zero. These state vectors are
mutually orthogonal for different helicities and are normalized to
unity if
\begin{equation}
\langle\,\boldsymbol{k}_{1a},\lambda_{1a},\boldsymbol{k}_{2a},\lambda_{2a}\,|\,\boldsymbol{k}_{1b},\lambda_{1b},\boldsymbol{k}_{2b},\lambda_{2b}\,\rangle=\delta_{\lambda_{1a}\lambda_{1b}}\delta_{\lambda_{2a}\lambda_{2b}}\,\delta^{3}(\boldsymbol{k}_{1a}-\boldsymbol{k}_{1b})\delta^{3}(\boldsymbol{k}_{2a}-\boldsymbol{k}_{2b}).\label{eq:4.2}
\end{equation}

Using similar methods to those employed in \citet{Hoffmann2017a},
these state vectors can be written on a basis of eigenvectors of total
four-momentum and total CM frame angular momentum as
\begin{multline}
|\,\psi(\boldsymbol{k}_{0},\boldsymbol{R},\lambda_{1},\lambda_{2})\,\rangle\\
=\int_{0}^{\infty}d\kappa\int d^{3}P\,\sum_{J=|\lambda_{1}-\lambda_{2}|}^{\infty}|\,\kappa,\boldsymbol{P},J,\lambda_{1}-\lambda_{2},\lambda_{1},\lambda_{2};f\,\rangle\,\frac{e^{-(\kappa-k_{0})^{2}/2\sigma_{p}^{2}}}{(\pi\sigma_{k}^{2})^{\frac{1}{4}}}\,e^{i\kappa R}\,\epsilon\sqrt{2J+1}\,e^{-\epsilon^{2}(J+\frac{1}{2})^{2}/2}\,\frac{e^{-\boldsymbol{P}^{2}/8\sigma_{p}^{2}}}{(4\pi\sigma_{k}^{2})^{\frac{3}{4}}}.\label{eq:4.3}
\end{multline}
As usual, we can ignore the dependence on $\boldsymbol{P}.$ Note
that, in this model, the average total three-momentum vanishes with
a small spread, so the frame is effectively a CM frame and we have
left off the CM labels from the helicities. We have changed quantum
numbers from total energy, $E,$ to $\kappa=E/2,$ which measures
the magnitude of momentum of either photon.

The behaviour of the CM-frame $PJM$ basis vectors under particle
exchange, $\mathcal{E},$ is found to be
\begin{equation}
U(\mathcal{E})\,|\,\kappa,\boldsymbol{0},J,\lambda_{1}-\lambda_{2},\lambda_{1},\lambda_{2}\,\rangle=|\,\kappa,\boldsymbol{0},J,\lambda_{1}-\lambda_{2},\lambda_{2},\lambda_{1}\,\rangle\,(-)^{J+\lambda_{1}-\lambda_{2}}.\label{eq:4.4}
\end{equation}
Then the exchange-symmetric state vector is
\begin{multline}
|\,\psi(\boldsymbol{k}_{0},\boldsymbol{R},\lambda_{1},\lambda_{2}),S\,\rangle=\int_{0}^{\infty}d\kappa\,\sum_{J=|\lambda_{1}-\lambda_{2}|}^{\infty}\frac{1}{\sqrt{2}}\{|\,\kappa,J,\lambda_{1}-\lambda_{2},\lambda_{1},\lambda_{2}\,\rangle+|\,\kappa,J,\lambda_{1}-\lambda_{2},\lambda_{2},\lambda_{1}\,\rangle\,(-)^{J+\lambda_{1}-\lambda_{2}}\}\times\\
\times\frac{e^{-(\kappa-k_{0})^{2}/2\sigma_{p}^{2}}}{(\pi\sigma_{k}^{2})^{\frac{1}{4}}}\,e^{i\kappa R}\,\epsilon\sqrt{2J+1}\,e^{-\epsilon^{2}(J+\frac{1}{2})^{2}/2},\label{eq:4.5}
\end{multline}
very nearly normalized to unity for these largely distinguishable
individual-particle wavepackets.

Note that we do not apply symmetrization to the separation eigenvectors,
as we want the probabilities of distinguishable measurements. In the
overlap between these and the state vectors of eq. (\ref{eq:4.5}),
only the case $M=\lambda_{1}-\lambda_{2}$ will contribute. We find
\begin{equation}
\langle\,r,J,\lambda_{1}-\lambda_{2},\lambda_{1},\lambda_{2},S\,|\,\psi(\boldsymbol{p},\boldsymbol{R},\lambda_{1},\lambda_{2}),S\,\rangle=\mathcal{A}(r,J,\lambda_{1},\lambda_{2})+\delta_{\lambda_{1}\lambda_{2}}\mathcal{A}_{e}(r,J,\lambda_{1},\lambda_{1}),\label{eq:4.6}
\end{equation}
where
\begin{equation}
\mathcal{A}(r,J,\lambda_{1},\lambda_{2})=\frac{1}{\sqrt{2}}\int_{0}^{\infty}d\kappa\,\sqrt{\frac{2}{\pi}}\,\kappa r\,j_{J-|\lambda_{1}-\lambda_{2}|}(\kappa r)\,\frac{e^{-(\kappa-k_{0})^{2}/2\sigma_{p}^{2}}}{(\pi\sigma_{p}^{2})^{\frac{1}{4}}}\,e^{i\kappa R}\,\epsilon\sqrt{2J+1}\,e^{-\epsilon^{2}(J+\frac{1}{2})^{2}/2}\label{eq:4.7}
\end{equation}
and
\begin{equation}
\mathcal{A}_{e}(r,J,\lambda_{1},\lambda_{1})=\frac{1}{\sqrt{2}}\int_{0}^{\infty}d\kappa\,\sqrt{\frac{2}{\pi}}\,\kappa r\,j_{J}(\kappa r)\,\frac{e^{-(\kappa-k_{0})^{2}/2\sigma_{p}^{2}}}{(\pi\sigma_{p}^{2})^{\frac{1}{4}}}\,e^{i\kappa R}\,(-)^{J}\epsilon\sqrt{2J+1}\,e^{-\epsilon^{2}(J+\frac{1}{2})^{2}/2}.\label{eq:4.8}
\end{equation}

We intend to use these results in the description of scattering experiments,
where we want the initial average separation, $R,$ to be much larger
than the spatial wavepacket width, $\sigma_{x},$ and the length scale
$1/p.$ In \citet{Hoffmann2017a}, we chose $R/\sigma_{x}=1/\sqrt{\epsilon},$
where $\epsilon=\sigma_{k}/p$ is the fractional average momentum
spread, chosen so small that $\sqrt{\epsilon}$ is also small. A suitable
choice is $\epsilon=0.001$ ($\sqrt{\epsilon}=0.03$), with sensible
results. Then
\begin{equation}
\frac{R}{\sigma_{x}}\gg1\quad\mathrm{and}\quad pR=\frac{1}{2\epsilon^{\frac{3}{2}}}\gg1.\label{eq:4.9}
\end{equation}
Exploring this regime will simplify our calculations. Exploring the
small $R$ regime would require more difficult calculations, and will
not be done here.

We will see shortly that the contributions to the amplitudes in eqs.
(\ref{eq:4.7},\ref{eq:4.8}) come mainly from $\kappa\cong p$ and
$r\cong R.$ In that region $\kappa r\gg1$ so we can use the asymptotic
approximation (\citet{Abramowitz1972})
\begin{equation}
\sqrt{\frac{2}{\pi}}\,\kappa r\,j_{\ell}(\kappa r)\rightarrow\sqrt{\frac{2}{\pi}}\,\sin(\kappa r-\ell\frac{\pi}{2}).\label{eq:4.10}
\end{equation}
In $\sin\varphi=\{\exp(+i\varphi)-\exp(-i\varphi)\}/2i,$ only the
$\exp(-i\varphi)$ term will contribute significantly, giving a peak
at $r=R.$ Then we can evaluate the $\kappa$ integrals to find
\begin{equation}
\mathcal{A}(r,J,\lambda_{1},\lambda_{2})\rightarrow\frac{1}{\sqrt{2}}\,e^{i(J-|\lambda_{1}-\lambda_{2}|-1)\pi/2}\,e^{-ik_{0}r}\,\frac{e^{-(r-R)^{2}/8\sigma_{r}^{2}}}{(4\pi\sigma_{r}^{2})^{\frac{1}{4}}}\,\epsilon\sqrt{2J+1}\,e^{-\epsilon^{2}(J+\frac{1}{2})^{2}/2}\label{eq:4.11}
\end{equation}
and
\begin{equation}
\mathcal{A}_{e}(r,J,\lambda_{1},\lambda_{2})\rightarrow\frac{1}{\sqrt{2}}\,e^{i(J-1)\pi/2}\,e^{-ik_{0}r}\,\frac{e^{-(r-R)^{2}/8\sigma_{r}^{2}}}{(4\pi\sigma_{r}^{2})^{\frac{1}{4}}}\,(-)^{J}\epsilon\sqrt{2J+1}\,e^{-\epsilon^{2}(J+\frac{1}{2})^{2}/2}.\label{eq:4.12}
\end{equation}

Then we want the total separation probability density, summed over
$J,$
\begin{multline}
\rho(r,\lambda_{1},\lambda_{2})=\sum_{J=|\lambda_{1}-\lambda_{2}|}^{\infty}\left|\mathcal{A}(r,J,\lambda_{1},\lambda_{2})+\delta_{\lambda_{1}\lambda_{2}}\,\mathcal{A}_{e}(r,J,\lambda_{1},\lambda_{1})\right|^{2}\\
=\frac{e^{-(r-R)^{2}/4\sigma_{r}^{2}}}{(4\pi\sigma_{r}^{2})^{\frac{1}{2}}}\sum_{J=|\lambda_{1}-\lambda_{2}|}^{\infty}\epsilon^{2}\,(2J+1)\,e^{-\epsilon^{2}(J+\frac{1}{2})^{2}}+\\
\quad+\delta_{\lambda_{1}\lambda_{2}}\,\frac{e^{-(r-R)^{2}/4\sigma_{r}^{2}}}{(4\pi\sigma_{r}^{2})^{\frac{1}{2}}}\sum_{J=|\lambda_{1}-\lambda_{2}|}^{\infty}(-)^{J}\epsilon^{2}\,(2J+1)\,e^{-\epsilon^{2}(J+\frac{1}{2})^{2}}.\label{eq:4.13}
\end{multline}
The difference between taking the $J$ sums from $|\lambda_{1}-\lambda_{2}|$
and from 0 is at most $\mathcal{O}(\epsilon^{2}).$ The first sum
then evaluates to unity while the last sum evaluates to $3.7\times10^{-7}$
for $\epsilon=0.001.$ So
\begin{equation}
\rho(r,\lambda_{1},\lambda_{2})\rightarrow\frac{e^{-(r-R)^{2}/4\sigma_{r}^{2}}}{(4\pi\sigma_{r}^{2})^{\frac{1}{2}}},\label{eq:4.14}
\end{equation}
a useful measure of the distribution in separation distance, normalized
to unity. We note that a different choice of $\ell$ would still lead
to a relative phase of unity between the two terms in eq. (\ref{eq:4.6}).

\section{Conclusions}

It was our aim to construct eigenvectors of photon separation distance,
consistent with quantum mechanics and special relativity. This was
done with a choice of free spherical waves in the matrix element between
momentum magnitude and separation distance eigenvectors. It was also
necessary to choose the ``orbital angular momentum'' index of those
free spherical waves. These can be considered postulates of the theory.
They were designed to make the photon case as close as possible to
the case of massive particles.

The construction was tested on a model problem meant to describe either
the initial or final state of a scattering experiment. We found that
this predicted a localized probability density in separation, although
in this case the spread of separation was very large.

The consequence is that one can now choose photon wavepackets with
an appropriate parameter, $R,$ to describe a scattering experiment,
knowing that the initial average separation of the wavepackets is
meaningfully given by $R.$

\bibliographystyle{apsrev4-1}

\end{document}